\PassOptionsToPackage{dvipsnames}{xcolor}
\pdfoutput=1
\documentclass[nonacm,sigconf]{acmart}
\usepackage{xcolor}
\usepackage{tikz}
\usepackage{caption}
\usepackage{subfigure}
\usepackage{subcaption}
\usepackage{multirow}
\usepackage{graphicx}
\usepackage{listings}

\newcommand\green[1]{\textcolor{ForestGreen}{#1}}

\lstdefinelanguage{Cypher}{
  morekeywords={
    MATCH, OPTIONAL, MERGE, CREATE, DELETE, DETACH, SET, REMOVE, FOREACH, WITH, 
    RETURN, WHERE, ORDER, BY, LIMIT, SKIP, DISTINCT, ASC, DESC, UNION, ALL, AS, 
    ON, AND, OR, XOR, NOT, EXISTS, CONTAINS, STARTS, ENDS, IN, SHORTESTPATH
  },
  sensitive=true,
  morestring=[b]",      
  morecomment=[l]{//},  
  morecomment=[s]{/*}{*/}, 
  keywordstyle=\bfseries\color{blue},
  commentstyle=\itshape\color{gray!70!black},
  stringstyle=\color{teal},
}

\AtBeginDocument{%
  }

\setcopyright{none}
\acmISBN{978-1-4503-XXXX-X/18/06}

\begin{document}

\title{What If: Causal Analysis with Graph Databases}
\author{Amedeo Pachera}
\affiliation{%
  \institution{Lyon1 University, CNRS Liris}
  \city{Lyon}
  \country{France}
}
\email{amedeo.pachera@univ-lyon1.fr}

\author{Mattia Palmiotto}
\affiliation{%
  \institution{Lyon1 University, CNRS Liris}
  \city{Lyon}
  \country{France}
}
\email{mattia.palmiotto@univ-lyon1.fr}

\author{Angela Bonifati}
\affiliation{%
  \institution{Lyon1 University, CNRS Liris \& IUF}
  \city{Lyon}
  \country{France}
}
\email{angela.bonifati@univ-lyon1.fr}

\author{Andrea Mauri}
\affiliation{%
  \institution{Lyon1 University, CNRS Liris}
  \city{Lyon}
  \country{France}
}
\email{andrea.mauri@univ-lyon1.fr}

\renewcommand{\shortauthors}{Pachera et al.}

\begin{abstract}
Graphs are expressive abstractions representing more effectively relationships in data and enabling data science tasks. They are also a widely adopted paradigm in causal inference focusing on causal directed acyclic graphs. Causal DAGs (Directed Acyclic Graphs) are manually curated by domain experts, but they are never validated, stored and integrated as data artifacts in a graph data management system.
In this paper, we delineate our vision to 
align these two paradigms, namely causal analysis and property graphs, the latter being the cornerstone of modern graph databases. 
To articulate this vision, a paradigm shift is required leading to rethinking property graph data models with hypernodes and structural equations, graph query semantics and query constructs, and the definition of graph views to account for causality operators. 
Moreover, several research problems and challenges arise aiming at automatically extracting causal models from the underlying graph observational data, aligning and integrating disparate causal graph models into unified ones along with their maintenance upon the changes in the underlying data. 
The above vision will allow to {\bf make graph databases aware of causal knowledge} and pave the way to data-driven personalized decision-making in several scientific fields.
\end{abstract}



\maketitle

\section{Introduction}




Causal Directed Acyclic Graphs (cDAGs) are a cornerstone of causal analysis, providing a rigorous framework to represent and analyze causal relationships between variables. By representing variables as nodes and causal connections as directed edges, cDAGs enable the visualization of complex systems, identification of confounders, mediators, and colliders, and the extraction of causal paths. These capabilities make cDAGs essential in diverse fields such as medicine, social sciences, and artificial intelligence, where understanding causal mechanisms is crucial. The advantages of cDAGs lie in their ability to disentangle direct and indirect effects, explicitly encode causal assumptions, and support unbiased estimation of causal effects when combined with sufficient adjustment sets.

Despite their utility, the construction of cDAGs remains a significant challenge. Traditionally, cDAGs are manually labeled and constructed by domain experts, a process that is not only time-consuming and costly but also prone to subjectivity. Different experts may interpret causal relationships differently, leading to variations in the structure of cDAGs across studies or datasets. Moreover, these discrepancies can manifest as differences in variables, causal paths, or even the inclusion of specific relationships, resulting in challenges when aligning and integrating multiple cDAGs from diverse data sources. These limitations reduce the scalability and consistency of causal analysis, especially in complex domains with evolving or heterogeneous data.





To address these challenges, we propose our vision of leveraging property graphs~\cite{AnglesABHRV17,2018Bonifati} as a foundational model for causal analysis. Property graphs, with their expressive power and flexibility, allow the encoding of multi-valued nodes and edges along with edge and node properties as key-value pairs. They represent the adopted data models for several open-source and commercial graph database tools (e.g. Neo4j~\cite{neo4j}, Amazon Neptune~\cite{amazonneptune}, Oracle PGX~\cite{oraclepgx}, SAP Hana Graph~\cite{saphana}, RedisGraph~\cite{redisgraph}, Sparksee~\cite{sparksee}, Kuzu~\cite{kuzu} etc.) and have recently been standardized with the appearance of graph query languages, such as GQL and SQL/PGQ. 
Unlike simple graphs, property graphs cater for the representation of both nodes and relationships with rich attributes, enabling a natural mapping of causal variables and relationships to nodes and edges. This representation can be further used to extract causal paths and align disparate data sources with greater consistency and efficiency. 
\begin{figure}[t!]
    \centering
    \includegraphics[width=0.9\linewidth]{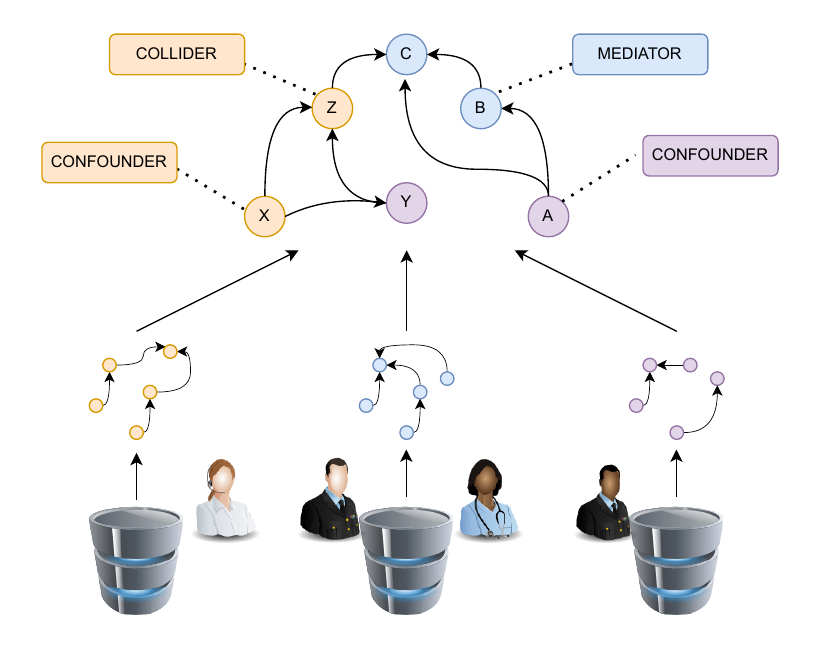}
    \caption{Path extraction from causal DAGs and graph databases.}
    \label{fig:intro}
\end{figure}
Another advantage of property graphs is their compatibility with concrete graph query languages, such as openCypher, GQL or SQL/PGQ \cite{DBLP:conf/sigmod/DeutschFGHLLLMM22, Soonbo2024view}. These languages enable graph pattern matching operations, retrieving and outputting results under the form of tuples. Recently, high-level abstractions, such as path-based algebraic foundations of graph query languages, have been proposed to enable compositionality of graph queries, thus returning property graphs as results instead of plain relational tuples \cite{angles2024path}. 
However, currently, property graph and graph query languages are not causality aware.
\begin{example}
Figure~\ref{fig:intro} exemplifies a scenario in which several domain experts are working on observational data, which is interspersed in three graph databases. The state of affairs in causal analysis implies that each of these experts manually designs a causal DAG based on the manual exploration of the underlying dataset. The approach is not only manual but tailored to one causal DAG per studied problem and per dataset. In the ideal situation in which one can leverage the graph pattern matching capabilities of graph databases, causal analysis becomes generalizable and multiple alternative causal DAGs are automatically extracted from the data stored in the databases. Different paths (highlighted in color) can be merged and integrated into a unified causal DAG (top). \end{example}

The above scenario might turn pivotal in several applications, where graph databases are used today, such as fraud detection, financial and social networks, supply-chain, cybersecurity and epidemiology, to name a few \cite{DBLP:journals/cacm/SakrBVIAAAABBDV21}.
As an example, in epidemiology, contact networks model the interactions between individuals that facilitate disease transmission. Nodes represent individuals, while edges denote interactions like physical proximity or shared environments, with properties such as duration, frequency, and nature of contact. Whereas the analysis of such networks is performed in graph databases to compute chains of interactions between people, causal analysis on such graphs would allow researchers to identify how specific interventions, such as quarantine measures or vaccination campaigns, affect the spread of a disease. Different data sources containing contact networks from various areas of the world might provide different effects of interventions (as depicted in Fig.~\ref{fig:intro}). For example, examining causal relationships in different contact networks can determine whether isolating high-degree nodes (super-spreaders) causally reduces overall infection rates. This approach leads to more targeted interventions, enabling officials to make better public health policies.

In order to achieve the above vision, we leverage the underlying property graph data models and declarative formalisms to allow the precise extraction of all causal paths, identification of bias-inducing relationships, and inference on the extracted graph. By using graph queries, we can automate the discovery of confounder, mediator, and collider paths, as well as explore the causal structure across different datasets. 
Existing works on using causality to determine the responsibility of query results \cite{Roy2022} and the completeness of the variables in the causal DAG \cite{Youngmann2023} focus on the relational model and does not consider a graph-based approach. 
Furthermore, property graphs facilitate the alignment and integration of multiple cDAGs, overcoming discrepancies in variables and relationships by representing and reconciling them within a unified graph structure.

In summary, the research vision that we investigate with this line of work centers around the following fundamental question: 
\emph{How can we empower graph databases, property graph models and queries with causal analysis capabilities?}

In order to address the above question, in this paper we make the following main contributions: 

\begin{itemize}
\item  We propose and formalize a causal property graph model augmented with hypernodes representing possibly related causes and with meta-properties encoding conditional and interventional probabilities along with structural equations. 

\item We study the extensions needed in current graph query languages to support \textit{path-based semantics}~\cite{angles2024path} and \textit{do-calculus} constructs. The latters are meant to extract from property graph observational data different variants of causal paths, namely confounder paths, bias paths, mediator paths and collider paths. 

\item We present the required mechanisms to harmonize and integrate different causal graphs from different data sources into a unified causal graph, leveraging property graph view formalisms. 

\item We discuss the problem of maintaining the above alignment between causal DAGs and underlying property graph instances, using graph maintenance algorithms that need to be adapted to causality-enabled scenarios. 
\end{itemize}

The above vision will allow to {\bf make graph databases aware of causal knowledge} and pave the way to data-driven personalized decision-making in several scientific fields.


The paper is organized as follows: Section~\ref{sec:preliminaries} presents the preliminary concepts underlying causal models Section~\ref{sec:model} introduces the extension needed for property graphs, including how to extract causal DAGs and the definition of causal graph views. Section~\ref{sec:query} describes how to perform causal analysis using GQL/Cypher, also discussing possible language extensions to allow such analysis. Section~\ref{sec:transport} discusses how graph data integration methods can be used to address the transportability problem of causal DAG. Section~\ref{sec:maintenance} describes how our models allow to easily maintain the causal DAG when the underlying data change. Finally, Section~\ref{sec:future} concludes and discusses future directions and challenges.

\section{CAUSAL MODELS}
\label{sec:preliminaries}
One of the most established theory of causality is represented by structural causal models (SCMs)~\cite{pearl2009causal}. SCMs consist of a causal graph and structural equations.
Causal graphs are a special class of Bayesian networks with the causal effect represented by edges, thus they inherit the well-defined conditional independence criteria~\cite{wang2008bayesstore}.
Formally, a causal graph $\mathcal{G}=(\mathcal{V},\mathcal{E})$ is a directed graph where $\mathcal{V}$ is the node set and $\mathcal{E}$ is the edge set. Each node in $\mathcal{V}$ represents a random variable while an edge $x \to y$ represents a causal effect between two variables $x,y \in \mathcal{V}$. Given an edge $x \to y$ we call $x$ the \textit{exposure} and $y$ the outcome. The node set $\mathcal{V}$ contains also all the observed and unobserved variables.
A \textit{directed path} is a sequence of directed edges that point to the same direction. We consider causal directed acyclic graphs (cDAGs), causal graphs that do not contain cycles~\cite{DechterP91}.
Figure~\ref{fig:intro} shows an example of cDAG with six variables. The path $A \to B \to C$ is a \textit{causal path}, representing a causal relation between the variables $A$ and $C$ through $B$ (called \textit{Mediator}). Moreover, variables $A$ and $X$ are \textit{confounders}, that is, they are the cause of two variables simultaneously. Finally, variables $Z$ and $C$ are called \textit{colliders}, because they have more than one incoming edge. In particular, $Z$ is a collider and a mediator at the same time but on different paths. 

\begin{figure}[t!]

    \centering
        \begin{tikzpicture}[->, node distance=2.5cm]
            \node[rectangle, rounded corners, draw] (x1) {Age};
            \node[rectangle, rounded corners, draw, right of=x1] (z1) {Income Level};
            \node[rectangle, rounded corners, draw, right of=z1] (y1) {Smoking};

            \draw (x1) -- (z1);
            \draw (z1) -- (y1);
        \end{tikzpicture}
    \\
    (a) Chain : Causal Path\vspace{5pt}
    
        \begin{tikzpicture}[->, node distance=2.5cm]
            \node[rectangle, rounded corners, draw] (x1) {Income Level};
            \node[rectangle, rounded corners, draw, right of=x1] (z1) {Age};
            \node[rectangle, rounded corners, draw, right of=z1] (y1) {Smoking};

            \draw[<-] (x1) -- (z1);
            \draw (z1) -- (y1);
        \end{tikzpicture}
    \\
    (b) Fork : Confounding Path\vspace{5pt}
    
        \begin{tikzpicture}[->, node distance=2.5cm]
            \node[rectangle, rounded corners, draw] (x1) {Age};
            \node[rectangle, rounded corners, draw, right of=x1] (z1) {Smoking};
            \node[rectangle, rounded corners, draw, right of=z1] (y1) {Income Level};
        
            \draw (x1) -- (z1);
            \draw[<-] (z1) -- (y1);
            
        \end{tikzpicture}
    \\
    (c) Collider : Collider Path
    
    \caption{DAGs for conditional independence.}
    \label{fig:dags}
\end{figure}
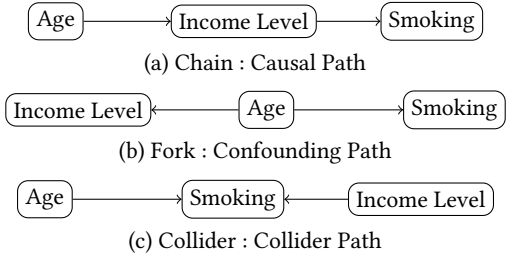

The conditional independence is guaranteed in causal graphs through \textit{dependency-separation} (d-separation)~\cite{pearl2009causal}. We say that two variables are d-separated if there is a \textit{blocked} node in the path between them where blocking means conditioning a variable on a set of other variables. For instance, Figure~\ref{fig:dags} shows three examples of DAGs. In the chain in Figure~\ref{fig:dags}(a), blocking the variable \textit{Income level} ($P(Income \; Level | Age)$) d-separates the variables \textit{Age} and \textit{Smoking}. Intuitively, once the income level is known, age has no more influence on the smoking habit. The same applies to the confounding path (Figure~\ref{fig:dags}(b)) : conditioning on \textit{Age} allows us to study the causal relationships between the income level and the smoking habit excluding the confounding bias.
On the contrary, if two variables are connected through a collider path, they are already d-separated (and thus causally independent). Conditioning the collider though, opens the path between the two variables. Conditioning on the node \textit{smoking} in Figure~\ref{fig:dags}(c) allows us to study the association between the age and the income level of an individual induced by the selection bias.

The second component of a Structural Causal Model consists of structural equations. Structural equations are non-parametric equations that quantify the causal effects between variables.
\begin{figure}[b!]
    \centering
    \begin{tikzpicture}[node distance=1.7cm, auto]

    \node (A) [draw, rectangle, rounded corners] {Age};
    \node (B) [draw, rectangle, rounded corners, below right of=A] {Smoking};
    \node (C) [draw, rectangle, rounded corners, below left of=A] {Income Level};

    \draw[->] (A) -- (B);
    \draw[<-] (B) -- (C);
    \draw[<-] (C) -- (A);

    \node (A2) [draw, rectangle, rounded corners] at (4.5,0) {Age};
    \node (B2) [draw, rectangle, rounded corners, below right of=A2] {Smoking};
    \node (C2) [draw, rectangle, rounded corners, below left of=A2] {do(100K)};

    \draw[->] (A2) -- (B2);
    \draw[<-] (B2) -- (C2);

\end{tikzpicture}
\\
   (a)\hspace{0.5\linewidth} (b)
    \caption{SCM without (a) and with (b) intervention.}
    \label{fig:SCMs}
\end{figure}
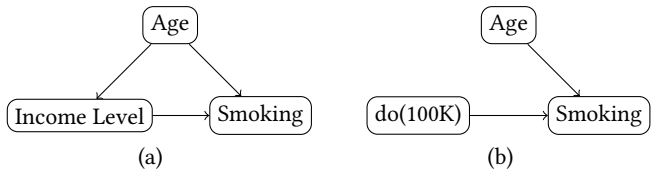

For instance, the structural equations for the SCM in Figure~\ref{fig:SCMs}(a) are:
$$ Age = f_{Age} (\epsilon_{age}) $$$$ Income\; Level=f_{Income\; Level}(Age,\epsilon_{Income\; Level})$$$$ Smoking=f_{Smoking}(Age,Income\; Level,\epsilon_{Smoking}),$$
where $\epsilon_x$ denotes the noise of the observed variable $x$. Note that the functions $f_x()$ are not invertible because they quantify the causal relationships between the LHS variable and the RHS one. Structural equations provide a quantitative way to represent \textit{intervention} on a variable in a DAG. The \textit{do-calculus}~\cite{pearl2009book} introduced the $do(x^{'})$ operator, which denotes the intervention of setting the value for a variable $x$ to $x^{'}$. Figure~\ref{fig:SCMs}(b) shows an example of intervention, where $do(100K)$ set the value for the variable \textit{Income Level} to \textit{100K}. The structural equation for \text{Income level} becomes indeed $Income\; Level = 100K$. With this intervention, we can formulate the \textit{interventional distribution} $$P(y | do(x^{'})) = f_y(x,\epsilon_x) $$$$ P(Smoking | do(100K)) = f_{Smoking}(Age,100K, \epsilon_{Smoking})$$
It is important to note that $P(y|do(x)) $ and $P(y|x)$ are not the same. This difference introduces a \textit{confounding bias}. In practice, verifying the absence of the bias translates into verifying that $P(y|do(x))=P(y|x)$.
In Figure~\ref{fig:SCMs}(a), with \textit{Age} being a confounder, the probability distribution $P(Smoking | Income\;Level)$ results from combining the causal effect $P(Smoking | do(Income\; Level))$ and the statistical association of the confounding path. To obtain an unbiased estimate, one needs to remove the confounding bias. This can be done via \textit{causal identification}, that is, blocking the confounder path. This operation requires to estimate the causal influence of the confounder on the other variables where the instances are homogeneous (i.e. adjustment).

\section{GRAPH MODEL}
\label{sec:model}

\begin{figure*}[t!]
    \centering
    \includegraphics[width=0.9 \linewidth]{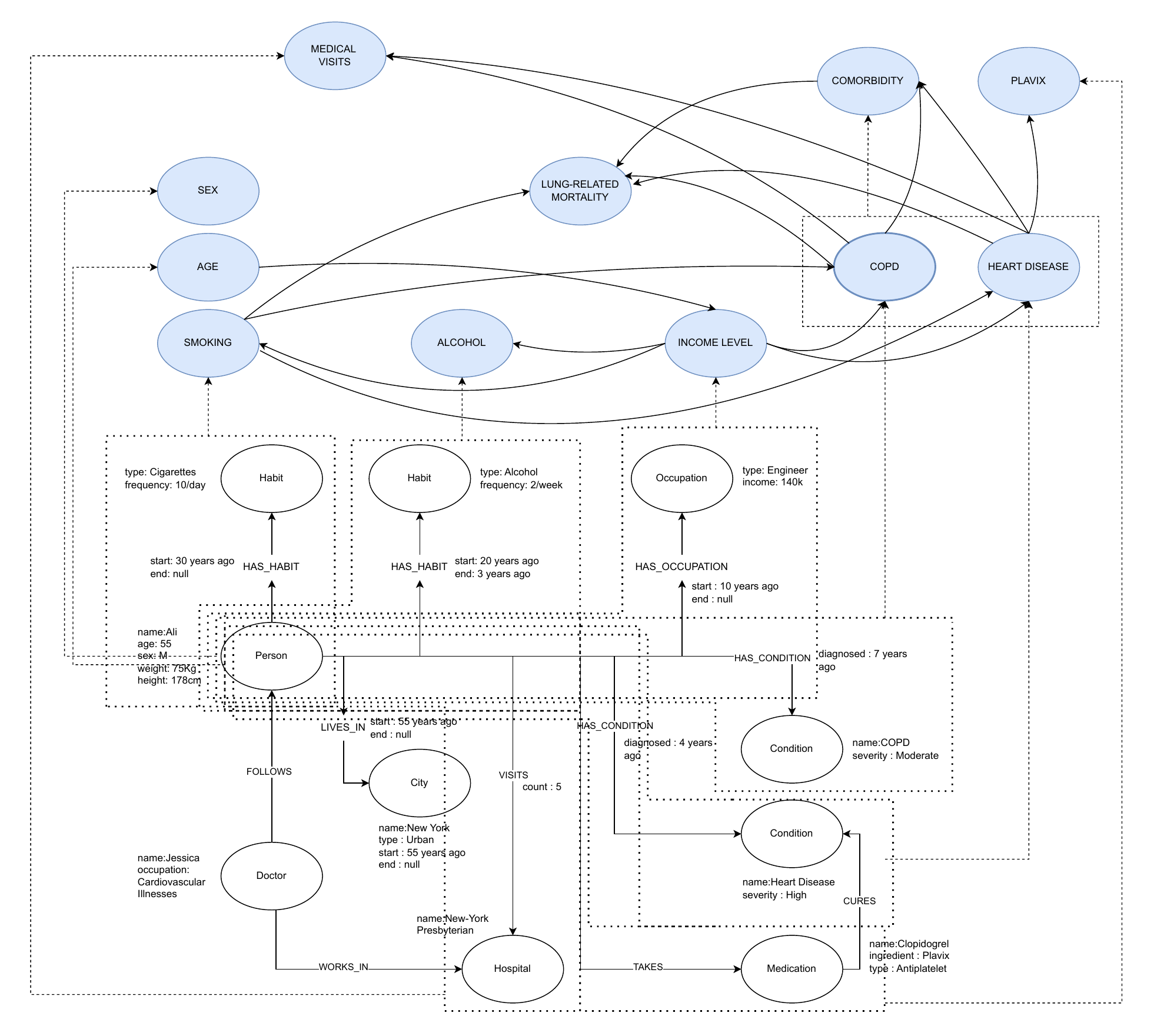}
    \caption{Example of a Causal Direct Acyclic Hypergraph model (variables of the causal DAG, while the property graph data is in white).}
    \label{fig:model}
\end{figure*}

Our vision leverages the expressiveness and flexibility of property graphs to streamline the extraction of Structural Causal Models, facilitate their maintenance, and enhance their interpretability. By embedding causal semantics directly into the graph structure, property graphs enable to perform causal analysis through queries using (or extending) existing graph data management methods, bridging the gap between raw data and actionable causal insights. Ideally, this approach should make the causal analysis process less tedious, as many of the activities that now requires ad-hoc scripting, can be executed in a declarative fashion.

Property graphs (PG) are multi-label directed graphs characterized by node and edge properties (i.e. metadata stored as key-value pairs)~\cite{2018Bonifati}. Unlike conventional DAGs, PGs naturally capture the rich context of causal systems, accommodating multi-faceted dependencies, metadata about variables and relationships, and evolving structures over time. These attributes make PGs uniquely suited for extracting cDAGs from complex data and maintaining them as systems evolve. Figure~\ref{fig:model} shows an example of a PG. It models a person named Ali who smokes and drinks some alcohol and has two diseases (\texttt{COPD} and \texttt{Heart disease}). 

To perform causal analysis, we need to extract causal variables from the PG and the relationships between them. Moreover, in order to apply interventions and perform causal analysis, we need to maintain a mapping between the PG instances and the causal variables in the cDAG. For this reason, we introduce a novel extension of the PG data model to cover hypergraphs. In particular, an hypervertex is a subgraph containing nodes and edges in the original property graph, but can be considered as a new node and can be linked to other existing vertices and new hypervertices. 

\textit{Causal Direct Acyclic Hypergraph (cDAH)}: 
Given a property graph $G$ with node set $V$ and edge set $E$, a Causal Direct Acyclic Hypergraph is a structure $(H, F, X, S, P, I, \gamma, \lambda, \delta, \eta, \mu, \nu)$, 
where $H$ is a finite set of hypervertices, $F$ is a finite set of edges with $F \cap E = \emptyset$, $X$ is a set of causal variables, $S$ is a set of structural equations, $P$ is a set of conditional probability distributions and $I$ is a set of interventional probability distributions.  $\gamma: V \to V \times E$ is a total function assigning every hypervertex a pair of vertex set and edge set such that the object sets $V$ and $E$ are disjoint and $\lambda : X \to H$ is an injective labeling function assigning one causal variable in $X$ to an hypernode in $H$. $\eta: X \to H$ is an injective function assigning a structural equation in $S$ to an hypernode in $H$. The bijective functions $\mu, \nu$ assign respectively a conditional probability distribution in $P$ and an interventional probability distribution in $I$ to each edge in $F$ as properties.

Notice that, for ease of exposition, we do not consider hyperegdes in the above definition. However, they could be easily embedded as further constructs identifying causal paths. 

Representing causal DAGs as property graphs leads to several advantages, namely the fact that metadata about the model and the actual data co-exist in the same data artifact and can be recorded and tracked. Moreover, this allows to exploit well know methods from graph data management (PG-Schema  \cite{AnglesBD0GHLLMM23}, PG-keys \cite{AnglesBDFHHLLLM21}, graph views~\cite{Soonbo2024view}, etc..), and, in particular, an expressive declarative language to perform the analysis (e.g., Cypher, GQL or SQL/PGQ). 



From the above definition, it is clear that hypernodes need to be extracted from the underlying property graphs. This can be done by defining property graph materialized views. 

\begin{figure}
    \centering
    \includegraphics[width=0.9\linewidth]{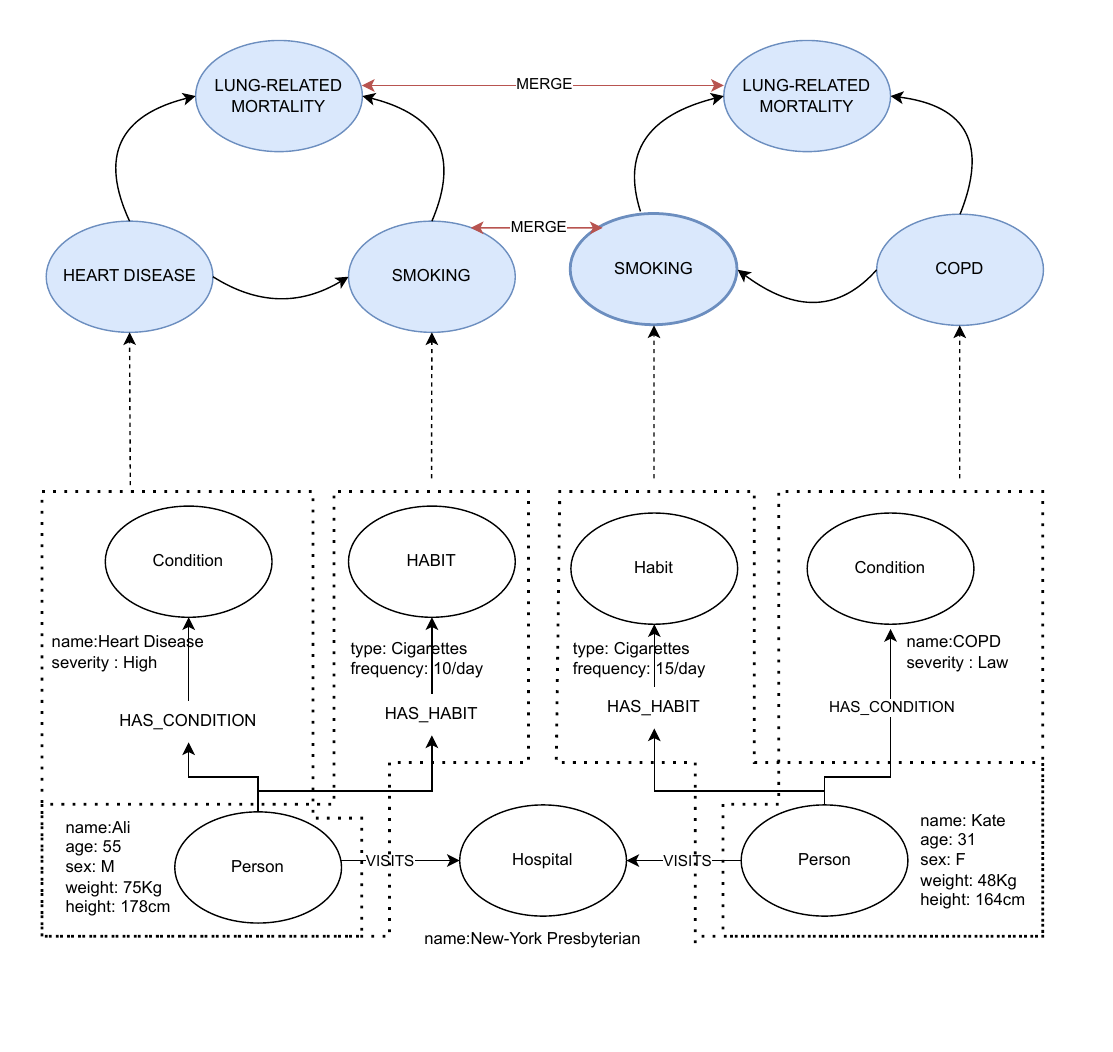}
    \caption{Example of merging of different nodes representing the same causal variable.}
    \label{fig:merge}
\end{figure}

For instance, the view definition in GQL as proposed in~\cite{Soonbo2024view} to extract the causal relation between the variables \textit{Smoking} and  \textit{Heart Disease} in the PG in Figure~\ref{fig:model} is:
\begin{lstlisting}
CREATE VIEW SmokingCausesHeartDisease AS (
|\color{blue}{CONSTRUCT}| (x:SMOKING)-->(y:HEART DISEASE)
MATCH (p:Person)-[:HAS_HABIT]->(h:Habit),
(p)-[:HAS_CONDITION]->(c:Condition)
WHERE c.name="Heart Disease" AND h.name="Cigarettes"
)
\end{lstlisting}

However, current graph DBMSs do not have support for views. A solution would be to adopt tuple generating dependencies~\cite{Soonbo2024view} or graph transformations~\cite{DBLP:journals/pvldb/BonifatiMR24}. An example of graph transformation merging \textit{Smoking} and  \textit{Heart Disease} in a causal DAG is the following Cypher query, where a new relationship is generated when a condition is met (the matched pattern):
\begin{lstlisting}
MATCH (p:Person)-[:HAS_HABIT]->(h:Habit),
(p)-[:HAS_CONDITION]->(c:Condition)
WHERE c.name = "Heart Disease" AND h.name="Cigarettes"
|\color{blue}{MERGE}| (x:SMOKING)-->(y:HEART DISEASE)
\end{lstlisting}
However, this query will produce a node (variable) for each matched path. Figure~\ref{fig:merge} shows an example of two variables (\textit{SMOKING} and \textit{LUNG-RELATED MORTALITY}) extracted twice from the property graph instance. 
To deal with duplicates, we should merge the generated nodes that have the same label. This is possible in concrete graph query languages but requires multiple complex queries. The first step of our vision consists of extending the GQL syntax with a new operator \green{\texttt{EXTRACT}}, that allows us to easily express the causal variable extraction by abstracting out the details of graph views and graph transformations:
\begin{lstlisting}
MATCH (p:Person)-[:HAS_HABIT]->(h:Habit),
(p)-[:HAS_CONDITION]->(c:Condition)
WHERE c.name = "Heart Disease" AND h.name="Cigarettes"
|\color{ForestGreen}{EXTRACT}| (x:SMOKING)-->(y:HEART DISEASE)
\end{lstlisting}
After extracting the variables from the property graph instance, the causal DAG is not yet complete as it needs to be enriched with the probability distributions to the edges as properties. In Section~\ref{sec:query}, we propose \textit{association queries} as tools for extracting these distributions directly from the property graph instance.

\section{CAUSAL PG QUERIES}
\label{sec:query}
Causal investigation consists of a three-level hierarchy of analysis: 1) Association, which serves as the starting point, identifying relationships between variables. However, association alone cannot answer causal queries but might guide hypothesis generation. It consists of posing questions like \emph{``What if I see..?''} or \emph{``How would seeing X change my belief in Y?''}. 2) Intervention, which is used for interventional queries, addressing questions like \emph{``What happens if we do X? Why?''} or \emph{``How can I make Y happen?''}. 3) Counterfactual, which is essential for providing insights into hypothetical alternate realities. It consists of questions like \emph{``What if I had done...? Why?''} or \emph{``What if X had not occurred?''}.
These investigations need to be paired with the structural analysis of causal models, which plays a crucial role in the broader process of causal discovery and causal inference. During the causal discovery phase, identifying confounder paths (and blocking them) prevents a biased interventional analysis because confounders can create spurious associations or hide true causal effects. The same occurs with colliders path because they can create spurious associations if incorrectly conditioned on.
In this section, we propose our vision about conducting causal analysis with cDAH, highlighting the advantages of relying on the property graph model. In particular, we identified a list of queries that correspond to a specific type of analysis and we show how to write them in Cypher. 

In our example, for a better understanding, we use \textbf{bold} for highlighting the part of the queries that refer to materialized views (e.g., the blue nodes in Figure~\ref{fig:model}) and the \green{green} color to highlight our proposed extensions to the query language. All the following graph queries are expressed in GQL, a standard graph query language published in May 2024~\cite{GQL2024}. Moreover, they can similarly be encoded in Cypher or SQL/PGQ~\cite{SQLPGQ2023}. Whenever procedural extensions are needed, we adopt Neo4j APOC procedures\footnote{https://neo4j.com/labs/apoc/}.

\subsection{Mediation and Confounding Path Queries}
Causal mediation analysis examines how an exposure influences an outcome through intermediary variables, known as mediators. This approach decomposes the total effect of an exposure into a direct effect, the impact of the exposure on the outcome not mediated by the intermediary, and an indirect effect, the portion of the effect that operates through the mediator. 
Identifying mediators between two variables $X$ and $Y$ can be achieved via this simple query:
\begin{lstlisting}
MATCH |\textbf{(a:X)}|-->|\textbf{(m)}|-->|\textbf{(b:Y)}|
RETURN a as exposure,
       m as Mediator, 
       labels(m) as MediatorVariable,
       b as outcome
\end{lstlisting}
The correct mediation analysis though, has to be conducted in absence of confounding effects between the variables. This implies to check for existing confounding paths between the investigated variables: 

\begin{lstlisting}
MATCH |\textbf{(a:X)}|-->|\textbf{(m)}|-->|\textbf{(b:Y)}|
WHERE NOT EXISTS {
MATCH  |\textbf{(a)}|<--|\textbf{(c)}|-->|\textbf{(b)}| }
|\color{blue}{AND}| NOT EXISTS {
MATCH |\textbf{(a)}|<--|\textbf{(c)}|-->|\textbf{(b)}| }
|\color{blue}{AND}| NOT EXISTS {
MATCH |\textbf{(m)}|<--|\textbf{(c)}|-->|\textbf{(b)}| }
RETURN a as Exposure,
       m as Mediator, 
       labels(m) as MediatorVariable,
       b as Outcome
\end{lstlisting}

However, with the current standard graph query languages (GQL and SQL/PGQ 1.0) it is not possible to return all the confounding paths. For this reason, we need to consider an extension of the query language with path operators, as proposed in~\cite{angles2024path}. 
This extension requires even more investigation if we consider (possibly) multiple mediators, and thus the pattern to match is \texttt{(a:X)-->*(m)-->(c:Y)},  where $*$ is the Kleene star operator that allows us to match patterns with 0 or more mediator nodes~\footnote{https://neo4j.com/docs/cypher-manual/current/patterns/reference/}.

The advantages of relying on PGs come from the possibility of accessing directly to the instances for further analysis. 

\begin{example} For example, if we want to know which is the mediator between the variable \texttt{SMOKING} and \texttt{COMORBIDITY} in the specific case of \texttt{Ali} in the graph in Figure~\ref{fig:model}, we can use the following query (avoiding for simplicity to check for confounding paths):
\begin{lstlisting}
MATCH (p:Person)-[BELONGS]->|\textbf{(a:SMOKING)}|,
|\textbf{(a)}|-->|\textbf{(m)}|-->|\textbf{(c:COMORBIDITY)}|
WHERE p.name="Ali"
RETURN m
WITH m
MATCH ALL instance=|\textbf{(m)}|<-[BELONGS]-(n1)
-[]|-|{*}(n2)|-|[BELONGS]->|\textbf{(m}|)
RETURN instance
\end{lstlisting}
Where \texttt{MATCH ALL} is the path operator that lists all the path that matches the given pattern.
This query first reaches the causal variable \texttt{SMOKING} from the node \texttt{Person}. Then identifies the mediator via pattern matching. Finally, it leverages the path algebra to return the subgraph that has generated the mediator variable. 
\end{example}
\subsection{Collider Path Query}
As we discussed in Section~\ref{sec:preliminaries}, colliders block the path between two variables. By conditioning on the collider, we can open the path and study the association between the variables. Identifying the colliders can be easily done via the following query:
\begin{lstlisting}
MATCH |\textbf{(a:X)}|-->|\textbf{(c)}|<--|\textbf{(b:Y)}|,
|\textbf{(c)}|-->{*}|\textbf{(ch)}|
RETURN c as Collider,
       labels(c) as ColldierVariable,
       ch as ChildCollider,
       labels(ch) as ChildColldierVariable
\end{lstlisting}
Using regular path queries \cite{2018Bonifati} has the advantage of allowing to list the children of the collider, since conditioning on one of them also opens the path~\cite{pearl2009causal}. Moreover, conditioning a variable means identifying it with a value. Similar to what we discussed with confounder path queries, we can leverage path algebra to directly access the instances of a collider path and select accordingly the right value to identify the collider random variable.

\begin{example} For example, in the graph in Figure~\ref{fig:model}, the path between \texttt{COPD} and \texttt{HEART DISEASE} through \texttt{COMORBIDITY} is blocked because \texttt{COMORBIDITY} is a collider. If we want to analyze the selection bias between having COPD and having heart disease we need to identify the variable \texttt{COMORBIDITY}. This way, we select a common outcome of the two variables and we can study the association between the two causes. To select a possible value for the variables, we can leverage graph queries and list all the possible values among the instances:

\begin{lstlisting}
MATCH ALL instance=|\textbf{(c:COMORBIDITY)}|<-[BELONGS]-(n1)
-[]|-|{*}(n2)|-|[BELONGS]->|\textbf{(c)}|)
RETURN DISTINCT(instance)
\end{lstlisting}
\end{example}

\subsection{Validation Query}
Validation queries check if the assumed causal structure is consistent with the data or domain knowledge. This type of query is useful when the cDAG is modeled by formalizing the causal assumptions and one wants to verify if the data reflects the assumed relationships. This includes validation of causal structures (e.g. \textit{``Do the data support the presence of a direct edge from X to Y in the causal graph?''}) or validation of d-separation (e.g. \textit{``Is X independent of Z when conditioning on Y?''}).
Thanks to cDAH, we can verify the statistical relationship between the variables by directly accessing the instances in the graph. However, computing the statistical relationship with the current GQL standard would require complex graph queries. For instance, we may use count queries to obtain the frequencies of different graph patterns, using the path algebra:
\begin{lstlisting}
MATCH ALL p=(a)-->(b),
RETURN COUNT(p) as PathCount
\end{lstlisting}
Then, we can chain multiple queries with the aggregation operators to derive the probability distributions. A better approach would consist of exploiting implementation libraries such as APOC for Neo4j to define procedures for computing the statistical relationships more efficiently. A possible example of a procedure defined using APOC for computing Conditional probabilities is: 

\begin{lstlisting}
CALL apoc.custom.asProcedure(
  |'|computeProbability|'|,
  |'|
  MATCH (x:|\$|xLabel)|-|[r]|-|>(y:|\$|yLabel)
  WITH y, x, count(r) AS pair_count
  MATCH (:`|\$|xLabel`)-[]->(y) 
  WITH y, x, pair_count, count(*) AS total_y
  RETURN y, x, pair_count * 1.0 |/| total_y AS probability
  |'|,
  |'|LIST OF MAP|'|,
  [
  [|'|xLabel|'|, |'|STRING|'|], 
  [|'|yLabel|'|, |'|STRING|'|]
  ]
);


//USAGE
CALL custom.computeProbability(|'|Person|'|, |'|Smoking|'|);
\end{lstlisting}

However, APOC procedures fall beyond the boundaries of a query language, while we envision to provide a declarative programming model for causal analysis. Hence, we propose to extend GQL/Cypher with a \texttt{PROBABILITY()} operator, as shown in the following example.

\begin{lstlisting}
MATCH |\textbf{(a:X)}||-|[r]|-|>|\textbf{(b:Y)}|,
WITH a,b
MATCH ALL instancesX=|\textbf{(a)}|<-[BELONGS]-(n1)
-[]|-|{*}(n2)|-|[BELONGS]->|\textbf{(a)}|
MATCH ALL instancesY=|\textbf{(b)}|<-[BELONGS]-(n1)
-[]|-|{*}(n2)|-|[BELONGS]->|\textbf{(b)}|
RETURN |\color{ForestGreen}{PROBABILITY(instanceY,instanceX)}|
\end{lstlisting}
\subsection{Association Query}
Association queries focus on identifying statistical relationships or dependencies between variables without making causal claims. This category of queries is the counterpart of validation queries and is useful when one wants to derive causal relationships from observed data. These queries include pairwise associations (e.g. \textit{``Is there a correlation between age and income in the population?''}) and group comparisons (e.g. \textit{``Do individuals with higher income levels smoke more cigarettes?''}).

The same approach is suitable for conditional associations, which consists of investigating associations while controlling for other variables. In this case, controlling the variable consists of filtering a specific graph pattern and then computing the statistical relations with the other variables. 
Finally, the property graph model allows us to investigate the more advanced types of associations such as multivariate associations (i.e. examining relationships among multiple variables simultaneously) by matching multiple paths, or temporal associations with time as the main 
component of the association. 

\subsection{Interventional Query}
Interventional queries focus on predicting or understanding the effects of specific interventions or manipulations. Traditionally, they involve tools like do-calculus~\cite{pearl1995causal} to simulate interventions. An example consists of investigating what will be the risk of lung-related mortality of a person who smokes 10 cigarettes a day. Thanks to our envisioned model, this question can be answered by traversing the path between the variables (\texttt{SMOKING} and \texttt{LUNG-RELATED MORTALITY} in Figure~\ref{fig:model}) to obtain the structural equations from the node properties and then applying the do-calculus ${do(cigarettes=10)}$:
\begin{lstlisting}
MATCH |\color{blue}{SHORTEST 1}| |\textbf{(a:SMOKING)}|-->{*}|\textbf{(m)}|,
|\textbf{(m)}|-->|\textbf{(b:LUNG-RELATED MORTALITY)}|
RETURN a.s_eq as s_s,
       m.s_eq as s_m,
       b.s_eq as s_l
\end{lstlisting}
where \texttt{SHORTEST 1} is the path algebra operator that matches the shortest path between two nodes~\cite{angles2024path}.
To solve the linear equations, we need to leverage again APOC procedures. An example of applying $do(cigarettes=10)$ is:
\begin{lstlisting}
MATCH |\textbf{(a:SMOKING)}|-->|\textbf{(b:COPD)}|
RETURN b.s_eq as s_b
WITH 10 as cigarettes
CALL apoc.math.evaluate(s_b, {SMOKING:cigarettes}) 
YIELD value
RETURN value
\end{lstlisting}
A better way to do it would consists in extending the GQL standard with an operator \green{\texttt{DO-CALCULUS([values],[equations])}} that evaluates the structural equations for an identified value.
An application to these queries consists of predicting missing values in the instances. 

\begin{example} For example, if the node \textit{Condition} with name \textit{COPD} in the graph in Figure~\ref{fig:model} is missing the severity level, we can apply do-calculus using the number of cigarettes smoked by \textit{Ali} (which we can access directly in the property graph):
\begin{lstlisting}
MATCH (p:Person)-[HAS_HABIT]->(h:Habit),
WHERE p.name="Ali" AND
      h.type="Cigarettes"
RETURN h.frequency as f
WITH f
MATCH |\color{blue}{SHORTEST 1}|  path=|\textbf{(a:SMOKING)}|-->|\textbf{(b:COPD)}|
RETURN |\color{ForestGreen}{DO-CALCULUS([SMOKING=f],[$b.s_{eq}$])}|
\end{lstlisting}
In this case, the query will return the missing value together with the noise of the random variable. Using property graphs as instances though, opens the possibility to also predict non-existing subgraphs.

Finally, interventional queries also pose hypothetical questions. If we want to predict for example how the severity of \textit{COPD} would change if \textit{Ali} reduces the number of cigarettes to 5, we can simply replace the last line with:

\begin{lstlisting}
RETURN |\color{ForestGreen}{DO-CALCULUS([SMOKING=5],[$b.s_{eq}$])}|
\end{lstlisting}
\end{example}
\subsection{Counterfactual Query}
Counterfactual queries focus on exploring ``what if'' scenarios about hypothetical alternatives. This implies do-calculus to force a variable to take a specific value regardless of its natural causes. Counterfactual analysis can be on an individual-level, for example asking \textit{``If patient A had not smoked, would they still have COPD?''}, or on a population-level, for example asking \textit{``If smoking rates had been reduced by 20\%, how much the lung-related mortality would have been decreased?''}. 

Counterfactual queries can be solved using the same principles as interventional queries because they both rely fundamentally on the framework of do-calculus. Interventional queries involve directly manipulating a variable ($do(X=x)$) to observe its effect on an outcome, while counterfactual queries extend this by conditioning on observed data to hypothesize alternate scenarios. The process of solving counterfactual queries often begins with using do-calculus to simulate the intervention within a causal model, followed by additional steps to account for observed evidence. This translates into computing the noise for the specific instance using the observed evidence.\\
\begin{example} For example, if we want to find out what the COPD severity of \textit{Ali} would have been if he had not smoked we can proceed by computing the noise of the variable \textit{COPD} from the observed data. This means that knowing that he smokes 10 cigarettes a day and the severity level of COPD is moderate, we can use the structural equation to derive the noise in his case. Then we apply do-calculus ${do(cigarettes=0)}$ to obtain the result. 
With our model, this translates into:

\begin{lstlisting}
MATCH (p:Person)-[HAS_HABIT]->(h:Habit),
(p)-[HAS_CONDITION]->(c:Condition)
WHERE p.name="Ali" AND
      h.type="Cigarettes" AND
      c.name="COPD"
RETURN h.frequency as f, c.severity as s
WITH f,s
MATCH |\color{blue}{SHORTEST 1}|  path=|\textbf{(a:SMOKING)}|-->|\textbf{(b:COPD)}|
RETURN |\color{ForestGreen}{DO-CALCULUS([SMOKING=f,COPD=s],[$b.s_{eq}$])}| as noise
WITH noise
MATCH |\color{blue}{SHORTEST 1}|  path=|\textbf{(a:SMOKING)}|-->|\textbf{(b:COPD)}|
RETURN |\color{ForestGreen}{DO-CALCULUS([SMOKING=0,$\epsilon_{COPD}$=noise],[$b.s_{eq}$])}|

\end{lstlisting}
\end{example}
\section{Causal DAGs Transportability}
\label{sec:transport}
 \begin{figure*}[htbp]
     \centering
     \includegraphics[width=0.8\linewidth]{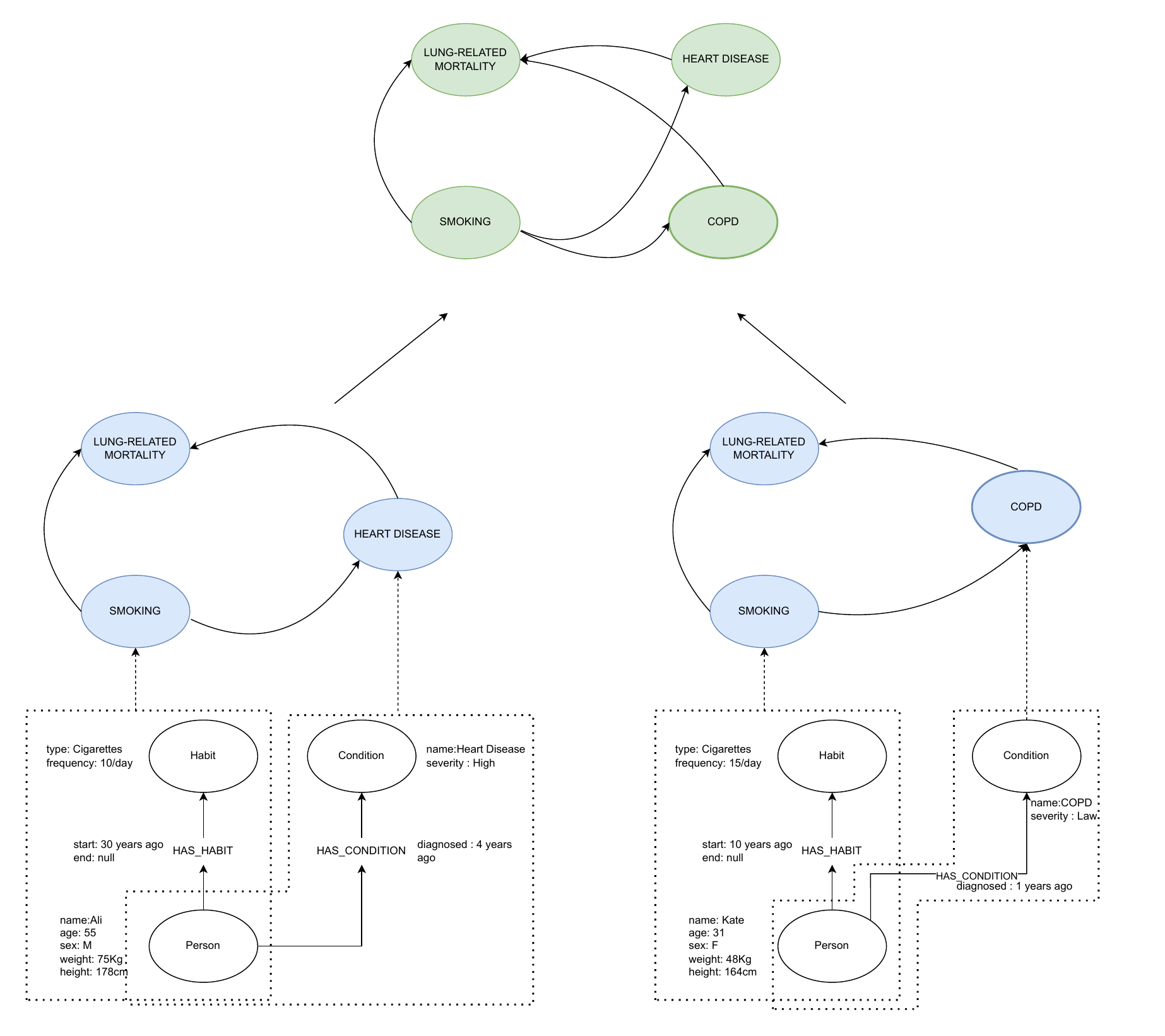}
     \caption{Example of cDAG alignment}
     \label{fig:alignment}
 \end{figure*}

Transportability is a concept in causal inference that addresses the challenge of applying causal knowledge derived in one context to another~\cite{pearl2011transportability}. It involves assessing whether the causal relationships established in one dataset or environment remain valid when applied to a different dataset or context. This process is particularly important in settings where data sources differ in population characteristics, experimental conditions, or contextual variables. By explicitly considering transportability, we can identify and adjust for confounding factors or contextual differences, ensuring robust and reliable causal inferences.
When working with multiple causal directed acyclic graphs (cDAGs) derived from different datasets, transportability becomes a foundational consideration. Each cDAG captures causal relationships within its specific context, but combining them requires addressing differences and inconsistencies between the graphs.
To solve this challenge we propose to adopt a Global-as-View (GAV) approach for integrating these individual cDAGs into a global schema~\cite{lenzerini2002lav}. 
By using GAV, we define the global schema as a set of mappings derived from the local property graph schemas, ensuring that the integration directly reflects the underlying data sources. 
GAV mappings are inherently procedural in nature since we need to specify how to get the data of the global schema by means of queries over the sources. This characteristic makes query answering in GAV relatively straightforward. In our scenario, we need to integrate multiple cDAH into a global one by defining a mapping from the local causal graphs to the global one. To achieve this goal, a possibility is to leverage graph transformations proposed in~\cite{DBLP:journals/pvldb/BonifatiMR24}, to match the pattern on the local cDAH and generate the global variables.
However, dealing with causal graphs poses the additional challenge of combining the probability distributions. 
For combining probabilities, we weight the probability distributions from different local graphs using the following equation:
\begin{align}
    P^{XY}_{G_{\cup}} = \sum_{i=1}^{N} w^{XY}_{G_i}  P^{XY}_{G_i}\\
    w^{XY}_{G_i}=\frac{n^{XY}_{G_i}}{\sum_{j=1}^{N} n^{XY}_{G_j}}
\end{align}
where $P^{XY}_{G_{\cup}}$ is the probability distribution (conditional or interventional) related to the causal relationship $X \to Y$ for the union of all the local graphs $G_i$ and $w^{XY}_{G_i}$ is the weighting factor, computed as the number of occurrences of the relationship among the instances in the local graph $G_i$ over the total number of the occurrences. \\
This approach allows us to tackle three key scenarios:
(1) Merging Multiple cDAGs:
By creating a unified schema, we can merge causal information from multiple sources. The process involves aligning similar variables across graphs and resolving structural differences, ensuring that the resulting cDAG (cDAH in our model) captures the full range of causal relationships present in the individual datasets.
(2) Testing Transportability Across Sources:
A unified schema enables explicit testing of transportability between different data sources. This involves determining whether the causal relationships identified in one dataset remain valid in others and identifying adjustments needed to account for contextual differences, such as confounders or mediators.
(3) Comparing Causal Relations:
The global schema facilitates a systematic comparison of causal relationships across datasets. By highlighting shared patterns or discrepancies, we can uncover deeper insights into the phenomena being studied, while also identifying potential biases or dataset-specific artifacts.\\

\begin{example} Consider the example shown in Figure~\ref{fig:alignment}, where two datasets contain information about patients, their habits, and potential health conditions. Suppose the two cDAHs (depicted in blue) were mined independently. While they describe the same phenomena, there are notable differences between them due to variations in the datasets, such as population diversity or measurement methods. To integrate these graphs:
We define the graph transformations that generate the global cDAH (depicted in green). However, adjustments may be made, such as combining the probability distributions of the relationship $SMOKING \to LUNG-RELATED\; MORTALITY$ to account for differences between the populations represented in the datasets. Note that, by combining the two local graphs, the variable \texttt{SMOKING} becomes a confounder for \texttt{COPD} and \texttt{HEART DISEASE}, while \texttt{LUNG-RELATED MORTALITY} becomes a collider. This example shows that combining structural models from different sources can highlight potential relationships that were not discovered in the local instances.
\end{example}
The process of aligning cDAGs (or, equivalently, cDAHs) is not without challenges. First, merging graphs may introduce cycles, violating the acyclic property required of DAGs. Careful cycle detection and resolution methods are necessary to maintain validity. Second, structural equations associated with nodes must be combined meaningfully to reflect the integrated causal relationships. Finally, validation queries (as discussed in Section~\ref{sec:query}) should be included to ensure that the resulting cDAH is not only valid but also transportable across contexts.

\section{Causal DAG Maintenance}
\label{sec:maintenance}

cDAG maintenance is the task of automatically updating the causal DAG according to the changes in the underlying data. Currently, it is a very tedious process that entails manually rerunning all causal analysis on the new and updated data, as existing methods solely focus on the static case and disregard the dynamic case~\cite{pearl2009book,shimizu2006linear,spirtes2001causation}.

On the other hand, our integrated cDAH model allows to reutilize established data management techniques such as incremental view maintenance~\cite{Soonbo2024view,Zhuge98view} and reactive graph data management~\cite{ceri2024,ceri2024pg}. The idea is to use triggers to automatically manage the changes in the data and update the causal DAG.

Inspired by the recent proposal for triggers for property graph~\cite{ceri2024pg}, our approach will define a set of triggers for each hypernode in the cDAH. For instance, to maintain the node \texttt{SMOKING}, we need to define triggers that listen for changes on the corresponding edge pattern \texttt{Person -[HAS\_HABIT]-> Habit {type:"Cigarettes"}} (this is not supported in~\cite{ceri2024pg} and needs an extension).

Then, the condition part of the trigger consists of running the associated \textit{validation} query (as described in Section~\ref{sec:query}). If the cDAH is still valid (i.e., the probability computed by the \textit{validation} query is still above a certain threshold), the trigger needs to update the conditional and interventional probabilities on the edge, and the structural equations on the nodes.

On the other hand, if the cDAH is not longer valid, we need to recompute by running only the relevant \textit{association} queries, to see if new edges appear.

After the cDAH has been updated we need to check if it's still valid. If an edge has been changed, we need to check for cycles, but in general, we need to run the \textit{validation} queries to verify the new cDAH. It may happen that the new cDAH is not valid anymore if the changes in the data are big enough to require to add or removal nodes. In this case, the only solution is to recompute a new cDAH from scratch. 

\section{Conclusions and Future Directions}
\label{sec:future}

In this paper, we proposed a new graph data model and an extension of GQL/Cypher to integrate causal analysis into property graphs. We also discussed how data management approaches and concepts, such as graph queries, graph views, graph transformations and graph triggers, can be applied to perform various tasks related to causal inference.

This opens up exciting possibilities and challenges for future research. These innovations have the potential to reshape the landscape of causal inference and data analysis by leveraging advanced computational models and methodologies. Below, we outline further research opportunities and challenges:

\textbf{Ensuring the Scalability and Robustness of the Approach.}
As causal graphs grow in size and complexity, ensuring scalable solutions remains a pressing challenge. Efficient query optimization, incremental updates to causal property graphs, and real-time graph data integration are essential for maintaining performance. Additionally, robustness to noisy or incomplete data must be addressed to make the data model applicable in diverse real-world scenarios.

\textbf{Designing a DSL for Causal Graph Analysis.}
A domain-specific language (DSL) tailored for causal graph analysis can simplify interactions with causal models. Such a DSL would allow users to express causal queries declaratively, with constructs for identifying causal paths, defining interventions, and performing counterfactual reasoning. The translation of the causal queries (see Section~\ref{sec:query}) into this DSL could be automated, enabling even non-expert users to explore causal relationships.

\textbf{Human-Centered Causal Analysis.}
Interactive approaches can enhance user engagement in the causal analysis, for example by suggesting causal paths and identifying critical blocking variables dynamically in an interactive fashion. These could provide real-time feedback, such as visualizing the impact of interventions or suggesting alternative adjustments to refine causal queries. Interactive graph methods could also be leveraged for repairing causal variables or relationships~\cite{pachera2025} and for graph alignment\cite{qian2019systemer}.

\textbf{Leveraging Graph Neural Networks.}
Leveraging the property graph model makes our approach well-suited for graph neural networks (GNNs). GNNs could be utilized to compute probabilities, classify graph patterns, and even predict causal relations based on learned representations. However, challenges remain in adapting GNNs to incorporate causal principles, such as encoding d-separation and handling intervention effects.

\textbf{Transformer-Based Approaches for Causal Inference.}
Transformers and large language models (LLMs) could offer another promising avenue for causal analysis. Their ability to model complex relationships and context can be harnessed to infer causal relations, identify confounding variables, and predict the impact of interventions. Challenges include fine-tuning these models for causal reasoning and ensuring interpretability in their predictions.

\bibliographystyle{ACM-Reference-Format}
\bibliography{sample-base}

\end{document}